\immediate\write18{makeindex \jobname.nlo -s nomencl.ist -o \jobname.nls}
\documentclass[preprint,12pt]{elsarticle}

\makeatletter
\def\ps@pprintTitle{%
  \let\@oddhead\@empty
  \let\@evenhead\@empty
  \def\@oddfoot{\reset@font\hfil\thepage\hfil}
  \let\@evenfoot\@oddfoot
}
\makeatother

\usepackage{graphicx}
\usepackage{amssymb}
\usepackage{amsmath}
\usepackage{algorithm}
\usepackage{algorithmic}
\usepackage{graphicx}
\usepackage{rotating}
\usepackage{bbm}
\usepackage{multirow}
\usepackage{nomencl}
\makenomenclature
\usepackage{vruler}
\newtheorem{theorem}{Theorem}
\newdefinition{definition}{Definition}

\newtheorem{property}[theorem]{Property}

\newproof{proof}{Proof}

\newcommand{\argmin}{\operatornamewithlimits{argmin}}

\def\EE{\mathbb{E}}
\def\PP{\mathbb{P}}

\def\S{{\cal S}}

\def\st{{\rm s.t.}}
\def\RSA{\text{(RSA)}}
\def\ind{\mathbbm{1}}

\begin{document}
\begin{frontmatter}



\title{Chance Constrained Optimization for Targeted Internet Advertising}


\author[aol]{Antoine Deza}
\ead{deza@mcmaster.ca}
\author[dsb]{Kai Huang}
\ead{khuang@mcmaster.ca}
\author[cse]{Michael R. Metel}
\ead{metelm@mcmaster.ca}

\address[aol]{Advanced Optimization Laboratory, Department of Computing and Software, McMaster University, Hamilton, Ontario, Canada}
\address[dsb]{DeGroote School of Business, McMaster University, Hamilton, Ontario, Canada}
\address[cse]{School of Computational Science and Engineering, McMaster University, Hamilton, Ontario, Canada}

\begin{abstract}
We introduce a chance constrained optimization model for the fulfillment of guaranteed display Internet advertising campaigns. The proposed formulation for the allocation of display inventory takes into account the uncertainty of the supply of Internet viewers. We discuss and present theoretical and computational features of the model via Monte Carlo sampling and convex approximations. Theoretical upper and lower bounds are presented along with a numerical substantiation.
\end{abstract}


\end{frontmatter}


\section{Introduction}

Internet advertising has witnessed growth of 15\% in 2012, reaching \$36.6 billion in the United States~\cite{IAB13}. This field is markedly different from traditional media used by advertisers such as radio, television and newspaper. Information such as a user's profile, data input and past Internet activity allow marketers to display their advertisements to targeted audiences, resulting in an efficient use of their advertising budget and an improved experience for users.\\

Our work is interested in the planning of guaranteed display Internet advertising by an ad network, which acts as an intermediary between website publishers and advertisers. Advertisers purchase an advertising campaign from the ad network consisting of a guaranteed campaign goal, which is the number of ads to be displayed, and a set of viewer types, which describe who to show the campaign's ads to. Guaranteed display advertising campaigns are typically for brand awareness where the industry practice is for ad networks to maximize representativeness, which is accomplished by diplaying ads of each campaign as proportionally as possible to all targeted viewer types, see Yang et al.~\cite{yan10}.\\

Quadratic optimization programs for this problem have recently been developed by Turner~\cite{tur12} and Yang et al.~\cite{yan10}. In particular, Turner showed that performance metrics are maximized using a proposed allocation methodology assuming the viewer supply follows a certain distribution. Our work addresses the uncertainty in viewer supply using a chance constrained framework. Bharadwaj et al.~\cite{bha10} presented an extension to~\cite{yan10} tangential to our research, using a two-stage stochastic program with recourse, with the second stage selling or purchasing ads on the spot market if the realized supply is greater or less than expected. An alternative objective to spread ads across campaigns is to maximize entropy, see Tomlin~\cite{tom00}. We pursue the quadratic objective value approach motivated in part by the availability of advanced and efficient solvers.\\

We introduce the model in Section~\ref{sec:mod} and formulate the joint chance constrained optimization program to solve the ad network's problem. Section~\ref{sec:SA} discusses how lower and upper bounds can be found through sample approximations. In Section~\ref{sec:conv}, a convex approximation program is presented which can be used to find lower and upper bounds under different Internet viewer distribution assumptions. We conclude with Section~\ref{sec:comp}, which discusses the results of a computational substantiation of the introduced bounds. A nomenclature can be found at the end of the text.

\section{Chance Constrained Optimization Model}
\label{sec:mod}
\subsection{Definitions and Notation}
An online ad network is an aggregator of display ad slots, which it sells to advertisers in partnership with website publishers. For each guaranteed display advertising campaign, the ad network displays ads to a targeted set of viewers that fit certain criteria, such as by demographic or interest. Advertisers are able to choose their targeted set of viewers from the set $V$\nomenclature{$V$}{Set of viewer types.} of viewer types, which partitions the publishers' viewers by a predefined set of attributes. Namely, the supply of viewers is modeled as a $|V|$-dimensional random variable with mean vector $\mu$\nomenclature{$\mu$}{Mean vector of viewer type supply.} and covariance matrix $\Sigma$\nomenclature{$\Sigma$}{Covariance matrix of viewer type supply.}. Let $\S_v$\nomenclature{$\S_v$}{Supply of viewer type $v$.} denote the supply of incoming ad slots across all websites in the ad network loaded by individuals of viewer type $v \in V$, with $\mu_v$\nomenclature{$\mu_v$}{Mean supply from viewer type $v$.} and $\sigma_v$\nomenclature{$\sigma_v$}{Standard deviation of viewer type $v$'s supply.} being its mean and standard deviation respectively. Let $K$\nomenclature{$K$}{Set of advertising campaigns.} denote the set of advertising campaigns. For a campaign $k\in K$, the campaign goal $g_k$\nomenclature{$g_k$}{Campaign goal of campaign $k$.} is the number of ads to be displayed to viewers, which we assume is given. For research concerning optimal campaign goal sizes, see~\cite{ahmed14}. The subset of viewer types $V_k \subseteq V$\nomenclature{$V_k$}{Set of viewer types targeted by campaign $k$.} are the viewer types targeted by advertiser $k$.
The subset of campaigns $K_v\subseteq K$\nomenclature{$K_v$}{Set of campaigns which target viewer type $v$.} are the campaigns which target viewer type $v$.\\

This problem can be viewed as a stochastic transportation problem with each viewer type as a source with random supply and each advertising campaign as a sink with known demand. Each time a user loads a website affiliated with the ad network, a decision must be made as to which advertisement to display. This paper focuses on the high level planning stage at the beginning of each optimization time period, determining what proportion of ads from each viewer type to allocate to each applicable campaign. The decision variables of the ad network are $p_{vk}$\nomenclature{$p_{vk}$}{Proportion of viewer type $v$'s supply allocated to campaign $k \in K_v$.}, the proportion of each viewer type $v$'s supply allocated to each campaign $k \in K_v$. Another means of planning, especially when dealing with campaigns over short time periods, is by allocating ads to one minute time slots, whereby all visitors during each time period are shown the same ads, see~\cite{deane12}.

\subsection{Chance Constrained Optimization Program}
We introduce the optimization program to find the proportion allocations, $p_{vk}$, for all viewer types and targeting campaigns, with an explanation following.

\begin{alignat}{6}
\label{eq:CC}
\tag{CC}
&\min &&\sum_{k \in K} \frac{w_k}{|V_k|}\sum_{v \in V_k} (p_{vk}-q_k)^2 \\
&\st &&\sum_{k \in K_v} p_{vk}   && \leq 1              &&\hspace{25 pt} \forall v \in V\nonumber \\
&&&\PP(\sum_{v \in V_k} \S_v p_{vk} \geq g_k \text{, } \forall k \in K) &&\geq 1-\alpha     &&\hspace{25 pt}\nonumber \\
&&&q_k                                     &&=\frac{1}{|V_k|}\sum_{v \in V_k}p_{vk}      &&\hspace{25 pt}\forall k \in K\nonumber \\
&&&p_{vk}                                  &&\geq 0              && \hspace{25 pt}\forall k \in K, v \in V_k\nonumber
\end{alignat}

The first constraint ensures that no more than $100\%$ of a viewer type's supply is allocated. The fourth constraint ensures that proportions are non-negative. The second constraint captures the idea of guaranteed campaign fulfillment, which is interpreted as fulfillment with high probability. In particular, the second constraint ensures that all campaigns are fulfilled with a probability of at least $1-\alpha$, where $\alpha < 0.5$\nomenclature{$\alpha$}{Campaign un-fulfillment tolerance.} is the un-fulfillment tolerance.\\

Chance constrained programming has been used in many different fields such as finance~\cite{li95} and water resource management~\cite{wang14}. We model campaign fulfillment using a chance constraint for two reasons. The first is that the success of an advertising campaign is unlikely to change dramatically if $g_k$ ads or $(1-\epsilon)g_k$ ads are displayed for some small percentage $\epsilon$, whereas strictly requiring the former may significantly limit the number of advertising campaigns the ad network can accept. With the parameter $\alpha$, the ad network is able to balance advertiser  satisfaction with the total number of advertising campaigns executed. The second, more fundamental reason is that robust solutions are unlikely to exist without making strong assumptions on the underlying distribution of Internet viewers. $\PP(\cup_{k \in K}\{\sum_{v \in V_k} \S_v < g_k\})=0$ is a necessary condition for the existence of a robust solution. For distribution assumptions of viewer type supply where this condition does not hold, e.g., normal, Poisson, log-normal, there exists a minimal $\hat{\alpha}>0$ such that $\alpha\geq\hat{\alpha}$ for (\ref{eq:CC}) to be a feasible program.\\

The objective of the ad network is to maximize representativeness, which is achieved by allocating each campaign $k$'s ads across all $v \in V_k$ proportionally to the supply. Objectives of the following general form have been proposed for guaranteed advertising campaigns, see~\cite{tur12,yan10},
$$\min \sum_{k \in K} \sum_{v \in V_k} w_{vk}(p_{vk}-\frac{g_k}{\bar{\mu}_k})^2$$
where the $w_{vk}$'s are weights, $\bar{\mu}_k=\sum_{v \in V_k}\mu_v$ is the total expected supply from the viewer types targeted by campaign $k$, and $\frac{g_k}{\bar{\mu}_k}$ is the target proportion. The objective maximizes weighted representativeness of campaigns, assuming the ad network is constrained to fulfill campaigns in expectation. Given the chance constraint, an ideal feasible allocation is unknown a priori. To construct a convex objective function reflecting the representativeness, we propose to minimize the variance of each campaign's allocation proportions. The objective is then
$$\sum_{k \in K} \frac{w_k}{|V_k|}\sum_{v \in V_k} (p_{vk}-q_k)^2$$
where $q_k$ is the mean of the proportions allocated to campaign $k$ from viewer types in $V_k$, enforced in the third constraint, and the weights $w_k$\nomenclature{$w_k$}{Campaign $k$'s priority weighting.} represent the campaign's priority to the ad network.
For example, assume campaign $k$ targets 5 viewer types, and a feasible solution to (\ref{eq:CC}) includes the vector of proportions allocated to campaign $k$, $p_k=[0.2,0.3,0.1,0.4,0]$. The objective then attempts to set
$p_k=[\phi,\phi,\phi,\phi,\phi]$ for some unknown $\phi$, which would achieve perfect representativeness for campaign $k$. We decided to minimize variance, but other measures of dispersion could also have been used.\\

Joint chance constraint programs are in general difficult to solve due to their non-convexity and the numerical integration required to calculate  $\PP(\sum_{v \in V_k} \S_v p_{vk} \geq g_k \text{, } \forall k \in K)$, see Pagnoncelli et al.~\cite{pag09}. Sample Approximations provide theoretically well founded solution approaches where Monte Carlo sampling is used to generate approximate mixed integer programs, see~\cite{BirgeLouveaux2011BOOK}. In Section~\ref{sec:SA}, we present results enabling the construction of sample approximations which, when solved, achieve lower and upper bounds with high probability for (\ref{eq:CC}).

\section{Sample Approximations}
\label{sec:SA}
\subsection{SA Program}
The following program, (\ref{eq:SA}), is a finite approximation to (\ref{eq:CC}). For $i=1,...,N$, the $\S^i_v$'s are independently sampled supply scenarios. The binary variable $x_i=1$ enforces the fulfillment of all campaign goals in scenario $i$. The second, third and sixth constraints require that all campaign goals are satisfied in at least $\lceil(1-\xi)N\rceil$ scenarios, which approximates the joint chance constraint $\PP(\sum_{v \in V_k} \S_v p_{vk} \geq g_k \text{, } \forall k \in K) \geq 1-\xi$.
\begin{alignat}{6}
\label{eq:SA}
\tag{SA}
&\min &&\sum_{k \in K} \frac{w_k}{|V_k|}\sum_{v \in V_k} (p_{vk}-q_k)^2 \\
&\mbox{s.t. } &&\sum_{k \in K_v} p_{vk}   && \leq 1              &&\hspace{25 pt} \forall v \in V\nonumber \\
&&&\sum_{v \in V_k} \S^i_v p_{vk}           &&\geq x_ig_k         &&\hspace{25 pt}\forall k \in K, i=1,...,N\nonumber \\
&&&\sum_{i=1}^{N} x_i                      &&\geq \lceil(1-\xi)N\rceil       &&\hspace{25 pt}\nonumber \\
&&&q_k                                     &&=\frac{1}{|V_k|}\sum_{v \in V_k}p_{vk}      &&\hspace{25 pt}\forall k \in K\nonumber \\
&&&p_{vk}                                  &&\geq 0              && \hspace{25 pt}\forall k \in K, v \in V_k\nonumber \\
&&&x_i                                     &&\in \{0,1\}         &&\hspace{25 pt}\forall i=1,...,N \nonumber
\end{alignat}
We can obtain lower and upper bounds with high probability by solving (\ref{eq:SA}) with an appropriate choice for $N$\nomenclature{$N$}{Number of viewer type supply scenarios for (\ref{eq:SA}).} and $\xi$\nomenclature{$\xi$}{Campaign un-fulfillment tolerance for (\ref{eq:SA}).}.

\subsection{SA Lower Bound}
Assume (\ref{eq:CC}) is a feasible program with optimal objective value $z^{(\ref{eq:CC})*}$ and optimal solution $p_{vk}^{(\ref{eq:CC})*}$. Property~\ref{eq:lowerprop} determines the probability of $p_{vk}^{(\ref{eq:CC})*}$ being feasible in (\ref{eq:SA}), implying the optimal objective value of $(\ref{eq:SA})$, $z^{(\ref{eq:SA})*}\leq z^{(\ref{eq:CC})*}$. When (\ref{eq:CC}) is not feasible, $z^{(\ref{eq:SA})*}\leq z^{(\ref{eq:CC})*}$, using the convention that $z^{(\ref{eq:CC})*}=\infty$.

\begin{property}[Luedtke \& Ahmed~\cite{lue08}]\label{eq:lowerprop}
$\PP(z^{(\ref{eq:SA})*}\leq z^{(\ref{eq:CC})*})\geq \sum_{i=0}^{\lfloor\xi N \rfloor}{N\choose i}\alpha^i(1-\alpha)^{N-i}$.
\end{property}

\subsection{SA Upper Bound}
Property~\ref{eq:uppertheorem} requires that the objective is convex, the deterministic feasible region is convex and closed, and that the chance constraint mapping is closed and convex. Let  $\RSA$ be the robust version of (\ref{eq:SA}) with $\xi=0$. This implies all $x_i=1$ converting (\ref{eq:SA}) into a convex quadratic program. Property~\ref{eq:uppertheorem} gives the probability that $z^{\RSA*}\geq z^{(\ref{eq:CC})*}$.

\begin{property}[Calafiore \& Campi \cite{cal06}]\label{eq:uppertheorem}
\text{ }\\
$\PP(z^{\RSA*}\geq z^{(\ref{eq:CC})*})\geq 1-{N \choose |V_K|}(1-\alpha)^{N-|V_K|}$, where $|V_K|=\sum_{k=1}^{|K|}|V_k|$ is the number of decision variables.
\end{property}

\subsection{A Branching Scheme for the Branch-and-Bound Algorithm}
\label{subsec:heur}
In this subsection we discuss an aspect of the algorithm used to find sample approximation lower bounds, which enabled us to solve larger scale problems. Assuming we are in the midst of solving (\ref{eq:SA}), we must solve the following program, (\ref{eq:ReSA}), at node $m$ of the Branch-and-Bound algorithm.
\begin{alignat}{6}
\label{eq:ReSA}
\tag{$\text{SA}_m$}
&\min &&&&\sum_{k \in K} \frac{w_k}{|V_k|}\sum_{v \in V_k} (p_{vk}-q_k)^2 \\
&\mbox{s.t. } &&&&\sum_{k \in K_v} p_{vk}   &&\leq 1              &&\hspace{25 pt} \forall v \in V\nonumber\\
&&&&&\sum_{v \in V_k} \S^i_v p_{vk}           &&\geq x_ig_k         &&\hspace{25 pt}\forall k \in K, i=1,...,N\nonumber\\
&&&&&\sum_{i=1}^{N} x_i                      &&= \lceil(1-\xi)N\rceil\nonumber\\
&&&&&x^T\text{diag}(X^1_m)                     &&=\mathbf{1}^T\text{diag}(X^1_m)\nonumber\\
&&&&&x^T\text{diag}(X^0_m)                     &&=\mathbf{0}\nonumber\\
&&&&&q_k                                     &&=\frac{1}{|V_k|}\sum_{v \in V_k}p_{vk}      &&\hspace{25 pt}\forall k \in K\nonumber\\
&&&&&p_{vk}                                  &&\geq 0              && \hspace{25 pt}\forall k \in K, v \in V_k\nonumber\\
&&&&&x_i                                     &&\in [0,1]           && \hspace{25 pt}\forall i=1,...,N \nonumber
\end{alignat}
where $X^1_m$ and $X^0_m$ are binary vectors of length $N$ which indicate the $x_i$ set to one and zero at node $m$ of the branching tree. We use an equality in the constraint $\sum_{i=1}^{N} x_i= \lceil(1-\xi)N\rceil$, as for any integral optimal solution with $\sum_{i=1}^{N} x^*_i = \lceil(1-\xi)N\rceil +r$ for some $r\in Z_{>0}$, any $r$ $x_i^*$'s equal to 1 can be set to 0 with  $p^{(\text{SA}_m)*}$ remaining as the optimal solution.\\

After solving (\ref{eq:ReSA}), assume that $p^{(\text{SA}_m)*}$ is not feasible in (\ref{eq:SA}) and the optimal objective value, $z^{(\text{SA}_m)*}$ is less than the current upper bound. Thus,  we want to branch on one of the $x_i$ for $i \in \{l : X^1_m(l)=0, X^0_m(l)=0\}$.
We use the following heuristic which finds the scenario $j$ with the constraint which is the farthest from being satisfied on a percentage basis:
$j=\argmin\limits_{i:X^1_m(i)=0, X^0_m(i)=0}\min\limits_{k=1,...,|K|}\frac{\sum_{v \in V_k} \S^i_v p^{(\text{SA}_m)*}_{vk}}{g_k}\nonumber$. For the path with $x_j=1$, the branching tree can be effectively pruned as enforcing scenario $j$ will likely enforce other scenarios, and the path with $x_j=0$ will lead to near optimal solutions as $x_j$ is a promising candidate for one of the $\lfloor \xi N  \rfloor$ scenarios to discard.

\section{Convex Approximations}
\label{sec:conv}
In this section we present convex constraints which can replace the joint chance constraint in (\ref{eq:CC}) to achieve bounding convex programs efficiently solvable using an interior point method.
Let $\S_k$\nomenclature{$\S_k$}{Vector of the viewer types' supply which campaign $k$ targets.} be the vector of the viewer types' supply which campaign $k$ targets, with $\mu_k$\nomenclature{$\mu_k$}{Mean vector of the viewer types' supply which campaign $k$ targets.} being the $|V_k|$-dimensional mean vector and $\Sigma_k$\nomenclature{$\Sigma_k$}{Covariance matrix of the viewer types' supply which campaign $k$ targets.} being the $|V_k| \times |V_k|$ covariance matrix. In addition, let $p_k$\nomenclature{$p_k$}{Vector of proportions allocated to campaign $k$ from viewer types in $V_k$.} be the $|V_k|$-dimensional vector of proportions allocated to campaign $k$ from viewer types in $V_k$.

\subsection{Distribution-free Bounds}

This subsection assumes that we only have estimates for the first two moments with no knowledge of the underlying distribution. General methodologies for generating convex relaxations and restrictions have been developed by Ahmed \cite{ahmed2013} and  Nemirovski \& Shapiro \cite{nem06}, respectively.  We present lower and upper bounds based on classic probability inequalities.


\begin{property}[Distribution-free Lower Bound]\label{eq:genlow} Any feasible solution of (\ref{eq:CC}) satisfies the constraints $p_k^T\mu_k \geq (1-\alpha)g_k $ for $k \in K$.
\end{property}

\begin{proof}
Assume there exists a $k' \in K$ with $p_{k'}^T\mu_{k'} < (1-\alpha)g_{k'}$, then $\PP(p_k^T\S_k \geq g_k \text{ } \forall k \in K) \leq \PP(p_{k'}^T\S_{k'} \geq g_{k'})\leq \frac{p_{k'}^T\mu_{k'}}{g_{k'}} <1-\alpha$, where the second inequality follows from Markov's inequality.
\qed
\end{proof}

\begin{property}[Distribution-free Upper Bound]\label{eq:genhigh}
The constraints
\begin{alignat*}{6}
&g_k-p_k^T\mu_k+\sqrt{\frac{1-\alpha_k}{\alpha_k}}\sqrt{p_k^T\Sigma_k p_k}&&\leq 0 &&\hspace{25 pt} \forall k \in K
\end{alignat*}
where $\sum_{k \in K}\alpha_k = \alpha$, $\alpha_k > 0$, form a conservative approximation of $\PP(p_k^T\S_k \geq g_k \text{ } \forall k \in K) \geq 1- \alpha$.
\end{property}

\begin{proof}
Following the reasoning in \cite[Sec.~2]{nem06}, assume $\PP(p_k^T\S_k < g_k) \leq \alpha_k \text{ } \forall k \in K$, then
$\PP(\cup_{k \in K}\{p_k^T\S_k < g_k\}) \leq \sum_{k \in K} \PP(p_k^T\S_k < g_k) \leq \alpha$,
implying $\PP(p_k^T\S_k \geq g_k \text{ } \forall k \in K) \geq 1- \alpha$.
To show $\PP(p_k^T\S_k < g_k) \leq \alpha_k$, we use the one-sided Chebyshev inequality, $\PP(Y \leq \EE(Y) - b) \leq \frac{\text{Var}(Y)}{\text{Var}(Y) + b^2}$ for a random variable $Y$ and constant $b>0$,
\begin{alignat*}{6}
&\PP(p_k^T\S_k < g_k)&&\leq\PP(p_k^T\S_k \leq g_k)\\
&&&\leq \frac{p_k^T\Sigma_k p_k}{p_k^T\Sigma_k p_k + (p_k^T\mu_k-g_k)^2}\\
&&&\leq \frac{p_k^T\Sigma_k p_k}{p_k^T\Sigma_k p_k + \frac{1-\alpha_k}{\alpha_k}p_k^T\Sigma_k p_k}\\
&&&=\alpha_k.
\end{alignat*}
\noindent Therefore the conclusion holds.\qed
\end{proof}

\subsection{Bounds Assuming a Normal Distribution}

The normal distribution has been proposed in the literature for modeling viewer type supply, see~\cite{bha10}. This subsection presents convex approximations under the assumption that $\S_k$ follows a multivariate normal distribution, so that $p^T_k\S_k \sim N( p^T_k\mu_k, p^T_k\Sigma_k p_k)$. Let $F_k$ denote the cumulative distribution function of $p^T_k\S_k$.

\subsubsection{Normal Lower Bound}\label{sec:NormLB}
Requiring each campaign's probability of fulfillment to be at least $1-\alpha$ is necessary for feasibility in (\ref{eq:CC}), resulting in a convex relaxation. The chance constraint for each campaign is equivalent to a second-order cone constraint \cite{boy04},
\begin{alignat*}{6}
&\PP(p_k^T\S_k \geq g_k)=&&1-F_k(g_k) &&\geq 1-\alpha\\
&&&\text{ }F_k(g_k) &&\leq \alpha\\
&&&\text{ }g_k &&\leq F_k^{-1}(\alpha)\\
&&&\text{ }g_k &&\leq p^T_k\mu_k  + n_{\alpha} \sqrt{p^T_k\Sigma_k p_k},
\end{alignat*}
where $n_\alpha$ is the $\alpha$ percentile of a standard normal random variable.

\subsubsection{Normal Upper Bound}

As in Property \ref{eq:genhigh}, an upper bound can be found by requiring
$\PP(p_k^T\S_k \leq g_k) \leq \alpha_k$ for all $k \in K$. Assuming the viewer supply follows a normal distribution,
we can then use the constraints found in Subsection \ref{sec:NormLB} with $\alpha$ replaced by $\alpha_k$.

\subsection{Summary}
The above convex approximations can be obtained by solving the following convex approximation program (\ref{eq:CA}) with the proper choice of parameters $u_k$ and $h_k$ as summarized in Table~\ref{T1}.

\begin{alignat}{6}
\label{eq:CA}
\tag{CA}
&\min &&\sum_{k \in K} \frac{w_k}{|V_k|}\sum_{v \in V_k} (p_{vk}-q_k)^2  \\
&\mbox{s.t. } &&\sum_{k \in K_v} p_{vk}   &&\leq 1               &&\hspace{25 pt} \forall v \in V  \nonumber  \\
&&&p^T_k \mu_k  - u_{k} \sqrt{p^T_k\Sigma_k p_k} &&\geq h_k  &&\hspace{25 pt} \forall k \in K \nonumber  \\
&&&q_k                                     &&=\frac{1}{|V_k|}\sum_{v \in V_k}p_{vk}      &&\hspace{25 pt}\forall k \in K \nonumber  \\
&&&p_{vk}                                  &&\geq 0              && \hspace{25 pt}\forall k \in K, v \in V_k \nonumber
\end{alignat}

\begin{table}[htb]
\centerline{
\resizebox{0.65\textwidth}{!}{
\renewcommand{\arraystretch}{1.5}
\begin{tabular}{lcc}
  \hline
	\bf{Bound} &$\mathbf{u_k}$&$\mathbf{h_k}$\\
  \hline
  Distribution-free Lower Bound&$0$ &$(1-\alpha)g_k$\\
  \hline
  Distribution-free Upper Bound&$\sqrt{\frac{1-\alpha_k}{\alpha_k}}$&$g_k$\\
  \hline
  Normal Lower Bound&$-n_{\alpha}$&$g_k$\\
  \hline
  Normal Upper Bound&$-n_{\alpha_k}$&$g_k$\\
  \hline
\end{tabular}}}
\caption{Parameters in (\ref{eq:CA})} \label{T1}
\end{table}

\noindent
(\ref{eq:CA}) was solved using a primal-dual interior point algorithm. The algorithm used generalized logarithm barriers to solve for points on the central path, see~\cite[Ch.~11.6-11.8]{boy04}. To form the modified KKT conditions, the Jordan algebra for second-order cones is used to express the complementary slackness conditions of the second-order cone constraints, see~\cite{ali01}. The system of equations to solve for the Newton steps was simplified so that only a system involving the step of $p$, $\Delta p$, was required to be solved, with closed form expressions for the remaining dual variable steps in terms of their current value, $p$, and $\Delta p$. In order to maintain stability, a universal step size was found such that all variables remained feasible. The central path parameter $t$ is updated to equal a multiple of the reciprocal of the maximum error of the modified KKT conditions involving $t$. The algorithm quits when the maximum error of the modified KKT conditions is less than or equal to a small $\epsilon$ times the current objective function value. In order to find an initial feasible solution, we begin with an algebraic expression which spreads proportions relative to the campaign's goal size to expected targeted supply, which satisfies the first and fourth constraints of (\ref{eq:CA}). The Big M method is then used to find an initial solution feasible in the second set of constraints.

\subsection{Setting the $\alpha_k$'s}
\label{subsec:alg}
We now present an iterative method to calculate upper bounds.
When finding a distribution-free upper bound, (\ref{eq:CA}) is first solved with the $\alpha_k$'s set equal to $\frac{\alpha}{|K|}$, as proposed in \cite[Sec.~2]{nem06}.
Letting $p_k^*$ be the optimal solution, with optimal objective $z^{(\ref{eq:CA})*}$, the approximating constraint of Property \ref{eq:genhigh} can be rearranged as
$\alpha_k\geq\frac{p_k^{*T}\Sigma_k p_k^*}{p_k^{*T}\Sigma_k p_k^* + (g_k-p_k^{*T}\mu_k)^2}=\hat{\alpha}_k$. For any $k$ for which this constraint is not tight, we can set $\alpha_k=\hat{\alpha}_k$. As these tighter constraints are valid for (\ref{eq:CA}), resolving the optimization problem with the tighter constraints, (TCA),  will result in an objective value $z^{(TCA)*}=z^{(\ref{eq:CA})*}$. Assuming there was at least one constraint in (\ref{eq:CA})  which had slack, $\sum_k\alpha_k<\alpha$ in (TCA). The total slack $s=\alpha-\sum_k\alpha_k$ can be added evenly to all $\alpha_k$'s of the originally tight constraints in (\ref{eq:CA}). Solving this relaxation of (TCA), (RTCA), will result in an objective value $z^{(RTCA)*}\leq z^{(TCA)*}$. This process of redistributing slack among the $\alpha_k$'s is iterated until the improvement in the objective values becomes sufficiently small. The same process is used for the normal upper bound, where for all approximating constraints with slack, $\alpha_k$ is updated to equal $F_k(g_k)$. The algorithm for solving the distribution-free upper bound is presented below, with the necessary changes to solve for the normal upper bound in the comments.

\begin{algorithm}                      
\caption{Calculating the Distribution-Free Upper Bound}             
\label{alg1}                           
\begin{algorithmic}[1]                 
    \STATE $\alpha_k=\frac{\alpha}{|K|} \text{ }\forall k\in K$
    \STATE $u_k=\sqrt{\frac{1-\alpha_k}{\alpha_k}}\text{ }\forall k\in K$\COMMENT{$u_k=-n_{\alpha_k}$ for the normal upper bound.}
    \STATE $[z^*,p^*]=CA(u,g)$
    \STATE $z^*_{old}=\infty$
    \STATE $I=0^{|K|}$\COMMENT{Indicator vector with $k^{th}$ entry set to 1 when slack found in constraint associated with campaign $k$.}
    \WHILE{$z^*_{old}-z^*>0$}
    \STATE $s=0$\COMMENT{Stores total slack across all constraints.}
    \FOR {k=1:$|K|$}
    \IF{$\alpha_k>\hat{\alpha}_k$}
    \STATE $s=s+\alpha_k-\hat{\alpha}_k$\COMMENT{$\hat{\alpha}_k=F_k(g_k)$ for the normal upper bound.}
    \STATE $\alpha_k=\hat{\alpha}_k$
    \STATE $I_k=1$
    \ENDIF
    \ENDFOR
    \IF{$s>0 \text{ }\&\text{ }\sum_{j=1}^{|K|}I_j<|K|$}
    \FOR {k=1:$|K|$}
    \IF{$I_k=0$}
    \STATE $\alpha_k=\alpha_{k}+\frac{s}{|K|-\sum_{j=1}^{|K|}I_j}$
    \ENDIF
    \ENDFOR
    \ENDIF
    \STATE $z^*_{old}=z^*$
    \STATE $u_k=\sqrt{\frac{1-\alpha_k}{\alpha_k}}\text{ }\forall k\in K$\COMMENT{$u_k=-n_{\alpha_k}$ for the normal upper bound.}
    \STATE $[z^*,p^*]=CA(u,g)$
    \ENDWHILE
\end{algorithmic}
\end{algorithm}

\section{Computational Substantiation}
\label{sec:comp}

In this section we compare the solutions of the sample and convex approximations. All testing was conducted on a Windows 7 Home Premium 64-bit, Intel Core i5-2320 3GHz processor with 8 GB of RAM. All coding was done in Matlab R2012a interfaced with CPLEX 12.4 using YALMIP \cite{lof04} dated 13-Feb-2013. Ten random test problems were generated. For each test problem, the number of campaigns and viewer types were chosen randomly between [5,10] and [10,20]. Campaign targeting was achieved by generating a $|K| \times |V|$ matrix of Bernoulli random variables with $p=0.5$, with cell $(i,j)=1$ indicating that campaign $i$ targets viewer type $j$. If there was a campaign or viewer type which was not assigned at least one viewer type or campaign, then a random cell in the appropriate row or column was set to 1. A random vector of viewer type means were generated, with each mean following a uniform distribution between [1000,10000]. Given the mean, $\mu_v$, $\sigma^2_v$ was randomly generated uniformly within $[0.25,0.5] \times \mu_v$. A random correlation matrix was generated using the random Gram matrix approach \cite{hol91}. For each campaign, $g_k= U_{[0.5,0.75]}\sum_{v\in V_k}\frac{\mu_v}{|K_v|}$, where $U_{[0.5,0.75]}$ is uniform between $[0.5,0.75]$. $w_k=1$ for all campaigns.\\

The sample approximation parameters for each test problem were chosen so that the optimal solution is between the bounds with a probability of at least $99\%$. We tested all bounds sampling the viewer supply from a normal distribution. $\alpha=0.1$ for the first five test problems and $\alpha=0.05$ for the remaining five.\\

Let the probability of fulfillment (PF) equal $\PP(\sum_{v \in V_k} \S_v p_{vk} \geq g_k \text{, } \forall k \in K)$. This probability is estimated for all solutions by generating 100,000 supply scenarios. Indicator variables, $\ind_{\{\sum_{v \in V_k} \S^i_v p_{vk} \geq g_k \text{, } \forall k \in K\}}$, for each scenario $i$ were generated and treated as a Bernoulli sample. The $99\%$ one-sided confidence interval of the probability of fulfillment, $\hat{PF}$, was then estimated, $\PP(PF\geq \hat{PF})=0.99$.\\

Results for each test problem are displayed in the Appendix. Objective values were multiplied by $1000$ for readability. For the sample approximation bounds, the lower bound objective, the $99\%$ one-sided confidence interval of the lower bound solution's probability of fulfillment, the lower bound computation time using the branching heuristic of Subsection \ref{subsec:heur}, the lower bound computation time solving directly with CPLEX, the upper bound objective, the $99\%$ one-sided confidence interval of the upper bound solution's probability of fulfillment, and the upper bound computation time are presented from left to right in Table \ref{T2} . When computing the lower bound directly with CPLEX, a time limit of $20*T_H$ was set, after which CPLEX would quit.\\

For the convex approximation bounds, the lower bound objective, the $99\%$ one-sided confidence interval of the lower bound solution's probability of fulfillment, and the lower bound computation time comprise columns 2-4 of Tables \ref{T3} and \ref{T4}. With $\alpha_k=\frac{\alpha}{|K|}$, the upper bound objective, the $99\%$ one-sided confidence interval of the upper bound solution's probability of fulfillment, and the computation time follow in columns 5-7. Using the algorithm of Subsection \ref{subsec:alg}, the upper bound objective, the $99\%$ one-sided confidence interval of the upper bound solution's probability of fulfillment, and the computation time are displayed in columns 8-10.\\

The average optimality gap for the sample approximation bounds was $43\%$, with an average computation time of $320$ seconds using the heuristic, which significantly improved the computation time. The average optimality gap and computation time for the distribution-free bounds was $385\%$ and $0.21$ seconds using the algorithm. The average improvement of the upper bound using the algorithm was $16\%$. The average optimality gap and computation time for the normal bounds was $11\%$ and $0.15$ seconds using the algorithm. The average improvement of the upper bound using the algorithm was $4\%$.

\section{Conclusion and Future Research}

This paper presents a chance constrained optimization model for guaranteed displayed Internet advertising campaigns. A sample approximation program with a branching heuristic was developed, as well as convex approximations under normal and distribution-free viewer supply assumptions, with an iterative method for improved optimality of feasible solutions.
Log-normal and Poisson distributions have also been proposed to model viewer supply, see~\cite{bha10,gir12}. Convex approximations under these assumptions is an area of potential future research.

\section*{Acknowledgements}
\noindent
This work was supported by grants from the Natural Sciences and Engineering Research Council of Canada, MITACS, and by the Canada Research Chairs program. The authors would like to thank the anonymous referees for their helpful comments.

\printnomenclature

\bibliographystyle{elsarticle-harv}
\bibliography{adpaper-references}

\section*{Appendix}

\begin{table}[htb]
\centerline{
\resizebox{0.84\textwidth}{!}{
\begin{tabular}{rrrrrrrrrr}
  \hline
	\multicolumn{10}{c}{\bf{Sample Approximation Bounds}} \\
  \hline
  \multirow{2}{*}{\#} && \multicolumn{4}{c}{\bf{LB}} && \multicolumn{3}{c}{\bf{UB}}\\  \cline{3-6}\cline{8-10}
  && \multicolumn{1}{c} {$z$} & \multicolumn{1}{c} {$\hat{PF}$} & \multicolumn{1}{c} {$T_H$ (s)} & \multicolumn{1}{c} {$T_C$ (s)} && \multicolumn{1}{c} {$z$} & \multicolumn{1}{c} {$\hat{PF}$} &\multicolumn{1}{c}{$T$ (s)}\\
  \hline
  $1$ && $0.08656$ & $0.67382$ & $394.12189$ & $7885.28954$  && $0.29815$ & $0.99883$ & $36.11891$\\
  $2$ && $38.47802$ & $0.83258$ & $271.55118$ & $5432.76618$ && $43.57778$ & $0.99931$ & $12.98950$\\
  $3$ && $1.39652$ & $0.72996$ & $118.50722$ & $2372.77228$  && $1.98864$ & $0.99448$ & $20.11220$\\
  $4$ && $147.26848$ & $0.78323$ & $642.50342$ & $5631.70639$&& $159.05476$ & $0.99384$ & $12.74514$\\
  $5$ && $1.55731$ & $0.76499$ & $228.41868$ & $4570.70080$  && $2.91692$ & $0.99728$ & $20.72935$\\
  $6$ && $100.38367$ & $0.76192$ & $139.77481$ & $1454.69702$&& $117.07098$ & $0.99687$ & $43.10787$\\
  $7$ && $209.21887$ & $0.78952$ & $375.03623$ & $3746.96343$&& $219.04089$ & $0.99844$ & $65.37173$\\
  $8$ && $29.24341$ & $0.83597$ & $69.03840$ & $1128.23085$  && $32.06909$ & $0.99838$ & $73.24367$\\
  $9$ && $0.00000$ & $0.99995$ & $0.45513$ & $1.50496$       && $0.00000$ & $0.99995$ & $36.52844$\\
  $10$&& $313.32316$ & $0.80525$ & $578.03043$ & $9679.22504$ && $338.78909$ & $0.99755$ & $57.58939$\\
  \hline
\end{tabular}}}
\caption{Results for Sample Approximation Bounds} \label{T2}
\end{table}

\begin{table}[htb]
\centerline{
\resizebox{\textwidth}{!}{
\begin{tabular}{rrrrrrrrrrrrr}
  \hline
	\multicolumn{13}{c}{\bf{Distribution-free Bounds}} \\
 \hline
 \multirow{2}{*}{\#} && \multicolumn{3}{c}{\bf{LB}} && \multicolumn{3}{c}{$\mathbf{UB_{\frac{\alpha}{|K|}}}$}&& \multicolumn{3}{c}{$\mathbf{UB_{ALG}}$}\\  \cline{3-5}\cline{7-9}\cline{11-13}
  &&\multicolumn{1}{c} {$z$}&\multicolumn{1}{c} {$\hat{PF}$}&\multicolumn{1}{c} {$T$ (s)}&&\multicolumn{1}{c} {$z$}&\multicolumn{1}{c} {$\hat{PF}$} & \multicolumn{1}{c} {$T$ (s)} && \multicolumn{1}{c} {$z$} & \multicolumn{1}{c} {$\hat{PF}$} & \multicolumn{1}{c} {$T$ (s)}\\
 \hline
$1$ && $0.02433$ & $0.10208$ & $0.04318$ && $1.56683$ & $0.99995$ & $0.13366$ && $0.70294$ & $0.99995$ & $0.33721$ \\
$2$ && $35.11906$ & $0.16493$ & $0.00708$ && $51.48740$ & $0.99995$ & $0.01638$ && $49.67440$ & $0.99995$ & $0.19283$ \\
$3$ && $1.07339$ & $0.11786$ & $0.00652$ && $4.90718$ & $0.99995$ & $0.02245$ && $3.27246$ & $0.99995$ & $0.27213$ \\
$4$ && $134.91058$ & $0.05268$ & $0.00575$ && $199.83478$ & $0.99995$ & $0.01840$ && $194.25912$ & $0.99995$ & $0.17045$ \\
$5$ && $0.91310$ & $0.17005$ & $0.00610$ && $9.19147$ & $0.99995$ & $0.01769$ && $4.96429$ & $0.99995$ & $0.24479$ \\
$6$ && $88.71204$ & $0.09186$ & $0.00600$ && $219.27627$ & $0.99995$ & $0.01562$ && $198.94794$ & $0.99995$ & $0.24656$ \\
$7$ && $200.41163$ & $0.01049$ & $0.00738$ && $276.35676$ & $0.99995$ & $0.02342$ && $266.59733$ & $0.99995$ & $0.20504$ \\
$8$ && $26.67377$ & $0.10453$ & $0.00721$ && $54.57600$ & $0.99995$ & $0.02209$ && $52.12026$ & $0.99995$ & $0.24255$ \\
$9$ && $0.00000$ & $0.99995$ & $0.01565$ && $0.00000$ & $0.99995$ & $0.01508$ && $0.00000$ & $0.99995$ & $0.02982$ \\
$10$ && $290.49479$ & $0.05637$ & $0.00673$ && $503.95350$ & $0.99995$ & $0.01890$ && $503.95350$ & $0.99995$ & $0.03657$\\
  \hline
\end{tabular}}}
\caption{Results for Distribution-free Bounds} \label{T3}
\end{table}

\begin{table}[htb]
\centerline{
\resizebox{\textwidth}{!}{
\begin{tabular}{rrrrrrrrrrrrr}
  \hline
	\multicolumn{13}{c}{\bf{Normal Bounds}} \\
 \hline
 \multirow{2}{*}{\#} && \multicolumn{3}{c}{\bf{LB}} && \multicolumn{3}{c}{$\mathbf{UB_{\frac{\alpha}{|K|}}}$}&& \multicolumn{3}{c}{$\mathbf{UB_{ALG}}$}\\  \cline{3-5}\cline{7-9}\cline{11-13}
  &&\multicolumn{1}{c} {$z$}&\multicolumn{1}{c} {$\hat{PF}$}&\multicolumn{1}{c} {$T$ (s)}&&\multicolumn{1}{c} {$z$}&\multicolumn{1}{c} {$\hat{PF}$} & \multicolumn{1}{c} {$T$ (s)} && \multicolumn{1}{c} {$z$} & \multicolumn{1}{c} {$\hat{PF}$} & \multicolumn{1}{c} {$T$ (s)}\\
 \hline
$1$ && $0.08488$ & $0.68915$ & $0.13207$ && $0.16233$ & $0.96111$ & $0.02831$ && $0.13225$ & $0.90751$ & $0.11698$ \\
$2$ && $37.80323$ & $0.72256$ & $0.04230$ && $39.48617$ & $0.93148$ & $0.01569$ && $39.27912$ & $0.91504$ & $0.04832$ \\
$3$ && $1.41510$ & $0.70816$ & $0.02301$ && $1.71973$ & $0.96048$ & $0.01950$ && $1.61711$ & $0.91542$ & $0.21837$ \\
$4$ && $144.25999$ & $0.61781$ & $0.01768$ && $150.74349$ & $0.92082$ & $0.01747$ && $150.16448$ & $0.90605$ & $0.05213$ \\
$5$ && $1.53287$ & $0.74523$ & $0.01855$ && $2.11183$ & $0.96274$ & $0.01825$ && $1.85374$ & $0.90628$ & $0.05421$ \\
$6$ && $100.78383$ & $0.80557$ & $0.01657$ && $107.56587$ & $0.96945$ & $0.02045$ && $106.72161$ & $0.96050$ & $0.06145$ \\
$7$ && $208.52184$ & $0.73829$ & $0.02060$ && $213.06374$ & $0.96324$ & $0.01995$ && $212.60627$ & $0.95321$ & $0.21879$ \\
$8$ && $29.15067$ & $0.83168$ & $0.02237$ && $30.56063$ & $0.97911$ & $0.02198$ && $30.08620$ & $0.95146$ & $0.07682$ \\
$9$ && $0.00000$ & $0.99995$ & $0.01484$ && $0.00000$ & $0.99995$ & $0.01449$ && $0.00000$ & $0.99995$ & $0.02862$ \\
$10$ && $311.30318$ & $0.73201$ & $0.02364$ && $323.19042$ & $0.95818$ & $0.02235$ && $323.18883$ & $0.95787$ & $0.24979$\\
  \hline
\end{tabular}}}
\caption{Results for Normal Bounds} \label{T4}
\end{table}

\end{document}